\documentclass[preprint]{aastex}
\usepackage{natbib}
\usepackage{booktabs}
\usepackage{threeparttable}

\usepackage{longtable}
\usepackage{tabularx}
\usepackage{lineno}

\begin{document}

\title{Quasi-periodic Accelerations of Energetic Particles during a Solar Flare}

\author{Dong~Li, and Wei~Chen}
\affil{Key Laboratory of Dark Matter and Space Astronomy, Purple Mountain Observatory, CAS, Nanjing 210023, PR China \\
     }
     \altaffiltext{}{Correspondence should be sent to: lidong@pmo.ac.cn}
\begin{abstract}
We report the observation of non-stationary Quasi-Periodic
Pulsations (QPPs) in high-energy particles during the impulsive
phase of an X4.8 flare on 2002 July 23 (SOL2002-07-23T00:35). The
X4.8 flare was simultaneously measured by the Reuven Ramaty High
Energy Solar Spectroscopic Imager, Nobeyama Radio Polarimeters, and
Nobeyama Radioheliograph. The quasi-period of $\sim$50$\pm$15~s,
determined by the wavelet transform, is detected in the $\gamma$-ray
line emission. Using the same method, a quasi-period of
$\sim$90$\pm$20~s is found in $\gamma$-ray continuum, hard X-ray
(HXR) and radio emissions during almost the same time. Our
observations suggest that the flare QPPs should be associated with
energetic ions and nonthermal electrons that quasi-periodically
accelerated by the repetitive magnetic reconnection. The different
quasi-periods between $\gamma$-ray line and continuum/HXR/radio
emissions indicate an apparent difference in acceleration or
propagation between energetic ions and nonthermal electrons of this
solar flare.
\end{abstract}


\section{Introduction}
Solar observations in $\gamma$-ray, hard X-ray (HXR), and radio
emissions have provided useful diagnostics for particle
accelerations of energetic ions and electrons on the solar system
\citep{Aschwanden02,Vilmer11}. Solar energetic particles, such as
nonthermal electrons and energetic ions, can interact with the solar
atmosphere and then produce radio (i.e. microwave), HXR,
$\gamma$-ray line and continuum emissions. Some energetic particles
may escape into the interplanetary space, generating the
low-frequency radio emission. The acceleration of electron beams in
solar flares has been established by detecting the radio, HXR, and
$\gamma$-ray continuum radiation, while the corresponding
acceleration of high-energy ions during large flares was prompted by
measuring the nuclear $\gamma$-ray line emission
\citep[e.g.][]{Vilmer12}. Spectroscopic observations in high-energy
channels have probed the behavior of sub-relativistic and
relativistic charged particles (i.e. ions and electrons) during
solar flares, such as the neutron-capture line at 2223~keV, the
positron-electron annihilation line at 511~keV, and the prompt
de-excitation $\gamma$-ray lines of heavy particles
\citep{Chupp83,Lin03,Share03,Smith03,Gan05,Murphy07,Chen20}. The
$\gamma$-ray line centered at 2223~keV is quite strong and extremely
narrow, it is generated when the thermalized neutron captured by the
ambient proton, regarded as the deuterium formation line. This
strong $\gamma$-ray line can be used as an indicator of $\gamma$-ray
flares, reflecting the radiation of nuclear reactions involving
flare-accelerated ions, and it has been well studied using the
imaging and spectroscopy observations, especially at the Reuven
Ramaty High Energy Solar Spectroscopic Imager
\citep[RHESSI;][]{Lin02} era, i.e., line shapes, line fluences, time
histories and spatially-resolved locations
\citep[e.g.][]{Holman03,Hurford03,Krucker03,Lin03,White03,Emslie04,Dauphin07,Chen12}.
While the quasi-periodic pulsations (QPPs) of this strong
$\gamma$-ray line during solar flares have been rarely reported.

QPPs are frequently observed as temporal intensity fluctuations
during solar/stellar flares \citep[see][for
reviews]{Van16,Zimovets21a}. They are often characterized by a
series of irregular but repetitive pulsations, termed as
`non-stationary QPPs' \citep[e.g.][]{Nakariakov19}. Generally, a
typical QPP event should have at least three successive peaks.
Because it is unnecessary to discuss the periodic behavior if there
are only one or two peaks, which might be just a coincidence, for
instance, the similar time interval between successive peaks
occurred by chance \citep{Nakariakov19,Li22a}. The flare-related
QPPs have been detected almost in the whole solar spectrum, i.e.,
radio/micowaves, white lights, H$\alpha$, ultraviolet/extreme
ultraviolet (UV/EUV), Ly$\alpha$, soft/hard X-rays (SXR/HXR) and
$\gamma$-rays
\citep[e.g.][]{Nakariakov10,Li20,Li20b,Clarke21,Hong21,Kashapova21,Zhao21,Karlick22,Li22b,Shi22}.
The observed quasi-periods are varied from sub-seconds to several
tens of minutes
\citep{Melnikov05,Ning17,Hayes20,Karlick21,Zimovets21b,Howard22,Ning22,Shen22,Zhou22}.
Here, the quasi-period refers to a slight variation of the dominant
period. The characteristic duration of all peaks in one QPP,
regarded as the period, is expected to be equal. However, the
durations in observations are mostly varied and irregular, and thus
regarded as `quasi-period' \citep[cf.][]{Nakariakov18}. The
quasi-period of a flare QPP is often associated with its generation
mechanism. Usually, the short-period QPP, such as $<$1~s, could be
related to the dynamic interaction of plasma waves with ambient
energetic particles in the complex magnetic structure
\citep{Tan10,Karlick22}, and the long-period QPPs, i.e., $>$1~s, are
frequently explained as magnetohydrodynamic (MHD) waves
\citep{Nakariakov20}. The flare QPP can also be driven by the
repetitive magnetic reconnection that can periodically accelerate
electrons and ions, and the quasi-period of reconnection process may
be either spontaneous or triggered
\citep{Nakariakov18,Yuan19,Clarke21,Li21,Karampelas22}. In a recent
review article \citep[e.g.][]{Zimovets21a}, a total of about fifteen
mechanisms/models were proposed to interpret flare QPPs. However, it
is still an open issue for the generation mechanism of flare QPPs,
mainly because that our observations cannot satisfy all the
necessary requirements to determine one mechanism that should be
responsible for a specific QPP event.

Thanks to the RHESSI HXR and $\gamma$-ray imaging and
spectroscopy observations, the X4.8 flare on 2002 July 23 had been
analyzed in many papers. For example, \cite{Lin03} presented an
overview of this flare observations, \cite{Hurford03} constructed
the first $\gamma$-ray maps, \cite{Krucker03} investigated the
movement of HXR sources, \cite{White03} compared the images between
radio and HXR emissions produced by high-energy electrons,
\cite{Holman03} analyzed the high-resolution HXR spectra,
\cite{Smith03} measured line profiles of de-excitation lines
generated by energetic ions, \cite{Murphy03} and \cite{Share03}
reported spectral observations of the neutron-capture line at
2223~keV and the positron-electron annihilation line at 511~keV,
respectively. However, the quasi-periodicity, especially the
$\gamma$-ray QPP, has not been reported during the X4.8 flare. In
this letter, we investigated non-stationary QPPs in $\gamma$-ray
line and continuum, HXR, and radio emissions during the flare
impulsive phase. Our observations revealed that the quasi-period
detected in $\gamma$-ray line emission was deviated from that
observed in $\gamma$-ray continuum, HXR, and radio emissions,
suggesting that the accelerated or propagated processes of energetic
ions and nonthermal electrons should be a bit different.

\section{Observations}
On 2002 July 23, an intense flare occurred in the active region of
NOAA 10039 near the solar east limb, i.e., S13E72. It was
simultaneously measured by the Geostationary Operational
Environmental Satellite (GOES), RHESSI \citep{Lin02}, Nobeyama Radio
Polarimeters (NoRP) and Nobeyama Radioheliograph
\citep[NoRH;][]{Nakajima04}, as shown in Figure~\ref{flux}.
Panel~(a) plots the GOES SXR flux at 1$-$8~{\AA} from 00:10~UT to
01:40~UT, which indicates an X4.8-class flare. The X4.8 flare began
at $\sim$00:18~UT and peaked at about 00:35~UT in the GOES flux, as
marked by the vertical orange lines. The GOES flare was accompanied
by a group of type III radio bursts, as shown by the context image
measured by Wind/Waves at the low-frequency range of
0.02$-$13.825~MHz.

RHESSI can provide SXR, HXR and $\gamma$-ray imaging spectroscopy of
solar flares from 3~keV to 17~MeV \citep{Lin02}.
Figure~\ref{flux}~(b) presents the full-disk light curves in HXR
emissions at 50$-$100~keV (black) and 100$-$300~keV (red), as well
as in $\gamma$-ray emissions at 300$-$500~keV (magenta),
700$-$1400~keV (green), and 2200$-$2300~keV (blue), respectively.
They have been normalized by their maximum values, and some light
curves have been shifted in height to display clearly in a same
window. Their time cadence is 4~s. To improve the signal-to-noise
(SN) ratio, the $\gamma$-ray line flux where the 2223~keV line
completely dominated, was integrated over a wide energy range of
100~keV, and thus we can obtain sufficient photon counts for QPP
test. The HXR maps were reconstructed by the RHESSI team and can be
directly downloaded from the RHESSI Image
Archive\footnote{https://hesperia.gsfc.nasa.gov/rhessi\_extras/flare\_images/2002/07/23/20020723\_0018\_0116/hsi\_20020723\_0018\_0116.html}.
Here, we used HXR maps with the CLEAN algorithm. On the other hand,
it is impossible to reconstruct the $\gamma$-ray map without the
help of RHESSI team. Herein, we used the centroid locations of
$\gamma$-ray line and continuum emissions obtained by
\cite{Hurford03}. We wanted to state that the RHESSI light curves
during about 00:26$-$00:35~UT were in a same
attenuator\footnote{http://hessi.ssl.berkeley.edu/hessidata/metadata/2002/07/22/hsi\_20020722\_235720\_rate.png}.
So, the HXR and $\gamma$-ray fluxes during our observations were not
affected by the RHESSI attenuator changes.

NoRH was designed to measure solar radio maps with a time cadence of
1~s at frequencies of 17~GHz and 34~GHz. However, there were some
data gap during our observations, resulting into some
discontinuities when integrated over light curves. Thus, the NoRH
light curve was unable to be used for QPP test. On the other hand,
NoRP could provide solar radio fluxes with an uniform resolution of
0.1~s at seven frequencies, such as 1~GHz, 2~GHz, 3.75~GHz, 9.4~GHz,
17~GHz, 35~GHz, and 80~GHz. Figure~\ref{flux}~(c) shows solar radio
fluxes normalized to their maximum values at three higher
frequencies recorded by NoRP during 00:25$-$00:33~UT. The X4.8 flare
was also observed by the Michelson Doppler Imager (MDI) on board the
Solar and Heliospheric Observatory (SOHO) and the Transiton Region
and Coronal Explorer (TRACE) at 195~{\AA}, which provided the
full-disk magnetograms and EUV maps, respectively.

\section{Methods and Results}
In Figure~\ref{flux}, the flare light curves in $\gamma$-rays, HXR,
and radio emissions are dominated by several successive peaks, which
appear to be irregular but repetitive. Thus, they could be
considered as a good candidate for non-stationary QPPs. It can be
seen that a sharp dip appears at roughly 00:30:20~UT, as indicated
by the red arrow. The sharp dip can be observed in $\gamma$-rays,
HXR, and radio fluxes, implying that some flare-accelerated
electrons/ions escaped from the Sun and propagate into the
interplanetary space. The synchronous low-frequency type III radio
burst observed by Wind/Waves in Figure~\ref{flux}~(a) confirmed the
presence of escaping electrons. We also note that the sharp dip of
$\gamma$-ray line flux is later than that of HXR and radio fluxes,
which is consistent with previous findings by comparing their
temporal profiles or the times of peak flux
\cite[e.g.][]{Lin03,Share03,Dauphin07,Vilmer11}, for instance, a
time delay of about 12~s was found between the $\gamma$-ray line
flux and the HXR flux at 150~keV \cite[cf.][]{Share03,Vilmer11}.

\begin{figure}
\centering
\includegraphics[width=\linewidth,clip=]{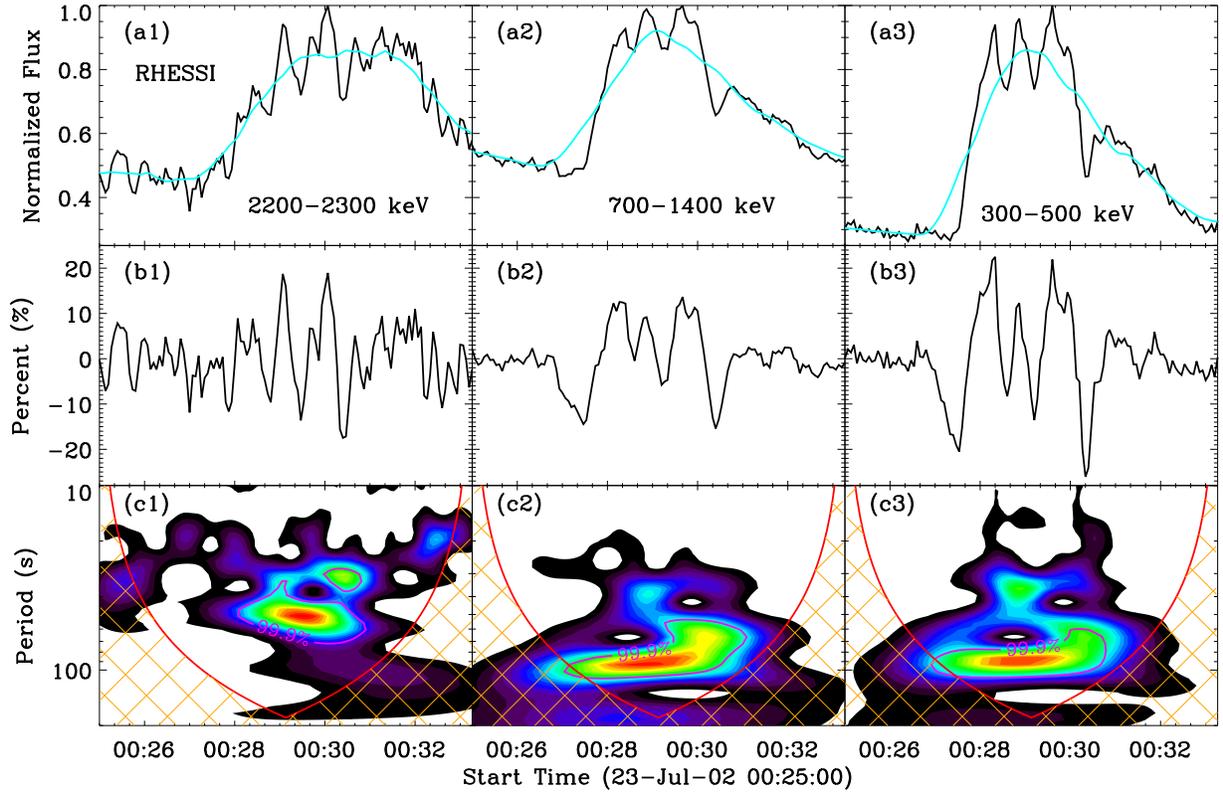}
\caption{(a1$-$a3) Flare fluxes normalized to their maximum values
in $\gamma$-ray line and continuum emissions observed by RHESSI
(black), the overlaid cyan lines are their slow-varying trends.
(b1$-$b3) Detrended light curves normalized to their maximum
slow-varying trends. (c1$-$c3) Morlet wavelet power spectra. The
magenta contours indicate a significance level of 99.9\%.
\label{wv1}}
\end{figure}

In order to take a closer look at the flare QPP in $\gamma$-ray
emissions, we performed a wavelet transform technique with the
mother function of `Morlet' \citep{Torrence98}. Figure~\ref{wv1}
presents the Morlet wavelet analysis results in $\gamma$-ray
emissions during the impulsive phase of the X4.8 flare. The upper
panels show flare light curves during $\sim$00:25$-$00:33~UT in
$\gamma$-ray line (a1) and continuum (a2 and a3) emissions,
respectively. Similar to previous studies
\citep[e.g.][]{Nakariakov10,Li20}, the raw light curves (black) were
firstly running average by smoothing 25~points (100-s window). Thus,
we obtained the slow-varying trends, as indicated by the overplotted
cyan lines. Next, the detrended fluxes in panels~(b1)$-$(b3) were
derived from the raw light curves after subtracting their
slow-varying trends. Both the raw and detrended light curves in
$\gamma$-ray continuum emissions at 700$-$1400~keV and 300$-$500~keV
appear at least three successive peaks from roughly 00:27~UT to
00:31~UT, while the successive peaks in the $\gamma$-ray line
emission seem to be more than three during the same time interval,
suggesting a short quasi-period. Moreover, the modulation depths of
these peaks, regarded as the ratio of the oscillatory amplitude to
their maximum slow-varying trend, are roughly 20\%$-$25\%, which is
consistent with what was found by \cite{Nakariakov10}. Finally, the
Morlet wavelet transform technique was applied to those detrended
light curves, as shown in panels~(c1)$-$(c3). These Morlet wavelet
power spectra reveal an enhanced power over a broad range of
periods, indicating the presence of QPPs at $\gamma$-ray levels. The
bulk of power spectra suggests that the flare QPP is characterized
by a dominant period within a large error. The dominant period is
determined from the center of enhanced power, while the error is
simply identified from the boundary of a 99.9\% significance level.
The quasi-period in $\gamma$-ray line emission is estimated to about
50$\pm$15~s, while that in $\gamma$-ray continuum emissions is
around 90$\pm$20~s. Obviously, the quasi-periods in $\gamma$-ray
line and continuum emissions are different.

\begin{figure}
\centering
\includegraphics[width=\linewidth,clip=]{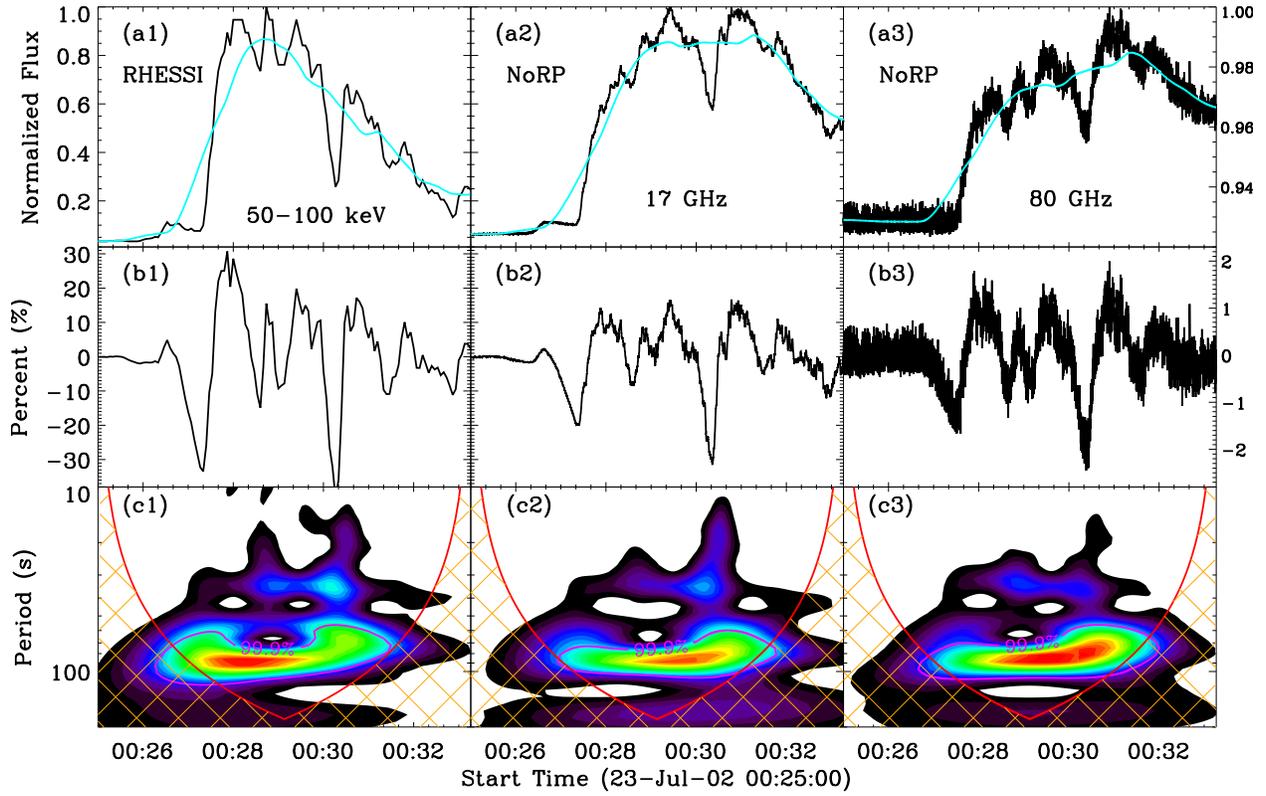}
\caption{Similar to Figure~\ref{wv1}, but for HXR and radio light
curves in the energy range of RHESSI~50$-$100~keV, and at
frequencies of NoRP~17~GHz and 80~GHz. \label{wv2}}
\end{figure}

Figure~\ref{wv2} presents Morlet wavelet analysis results in HXR and
radio emissions observed by RHESSI and NoRP, respectively.
Panels~(a1)$-$(a3) draw the raw light curves (black) and their
slow-varying trends (cyan). Here, the HXR light curve was smoothed
by 25~points, while the radio fluxes were smoothed by 1000~points,
and thus their smoothing window has a same temporal scale of 100~s,
same to that in $\gamma$-ray fluxes. Panels~(b1)$-$(b3) plot the
detrended light curves after removing the slow-varying trends from
their raw light curves. Similar to the $\gamma$-ray continuum
emission, both HXR and radio fluxes exhibit about three large-scale
peaks from roughly 00:27~UT to 00:31~UT. The modulation depths of
those peaks in HXR~50$-$100~keV and radio~17~GHz are estimated to be
20\%$-$30\%, roughly in agreement with those found in $\gamma$-ray
light curves. While the modulation depth in radio~80~GHz is much
smaller, i.e., only 1\%$-$2\%. On the other hand, the HXR and radio
fluxes appear a number of sub-peaks with very small amplitudes
during our observations, which is different from that in
$\gamma$-ray emissions, and thus it is beyond the scope of this
study. Panels~(c1)$-$(c3) show the Morlet wavelet power spectra,
revealing an enhanced power over a broad range of periods, which is
estimated to about 90$\pm$20~s. This quasi-period matches well with
that seen in $\gamma$-ray continuum emissions, but it is larger than
that found in the $\gamma$-ray line emission.

\begin{figure}
\centering
\includegraphics[width=\linewidth,clip=]{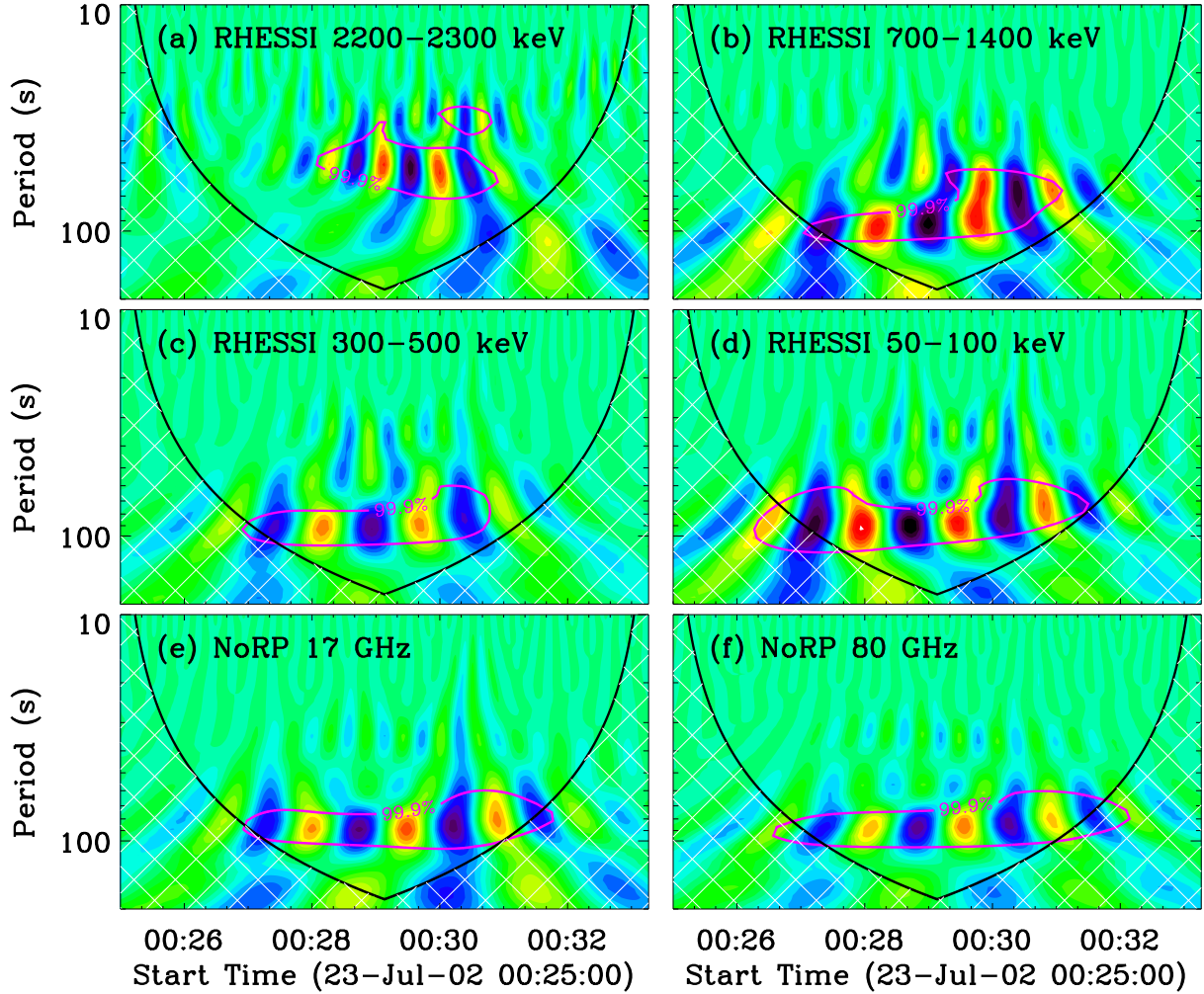}
\caption{Wavelet amplitude spectra corresponding to the detrended
time profiles seen in multiple wavelengths. The magenta contours
represent the 99.9\% significance level, and the color background
indicates the signal variations in time and periods. \label{wv3}}
\end{figure}

To look closely the spectral structure of the flare QPPs, we
constructed wavelet amplitude spectra \citep{Torrence98,Torrence99,Karlick20}. 
Figure~\ref{wv3} shows the amplitude wavelet spectra constructed from the detrended
time profiles in $\gamma$-ray line and continuum emissions, as well
as the HXR and radio emissions, the superimposed magenta contour in
each panel represents a significance level of 99.9\%. In panel~(a),
there is a periodicity with an average period of about 50~s inside
the 99.9\% significance level, confirming the presence of 50-s QPP
in the $\gamma$-ray line emission. In panels~(b)$-$(f), a
periodicity with an average period near 90~s is seen inside the
99.9\% significance level, suggesting the synchronous presence of a
dominant period at $\sim$90~s in $\gamma$-ray continuum, HXR and
radio emissions. The quasi-periods of those oscillations are all
manifested as bright and dark patches in the amplitude wavelet
spectra, similarly to what observed with the flare QPPs in radio,
EUV and X-ray emissions \citep[cf.][]{Karlick20}. On the other hand,
the average period in the $\gamma$-ray line is obvious smaller than
that seen in $\gamma$-ray continuum, HXR and radio emissions, in
agreement with the quasi-period seen in the wavelet power spectra.

\begin{figure}
\centering
\includegraphics[width=\linewidth,clip=]{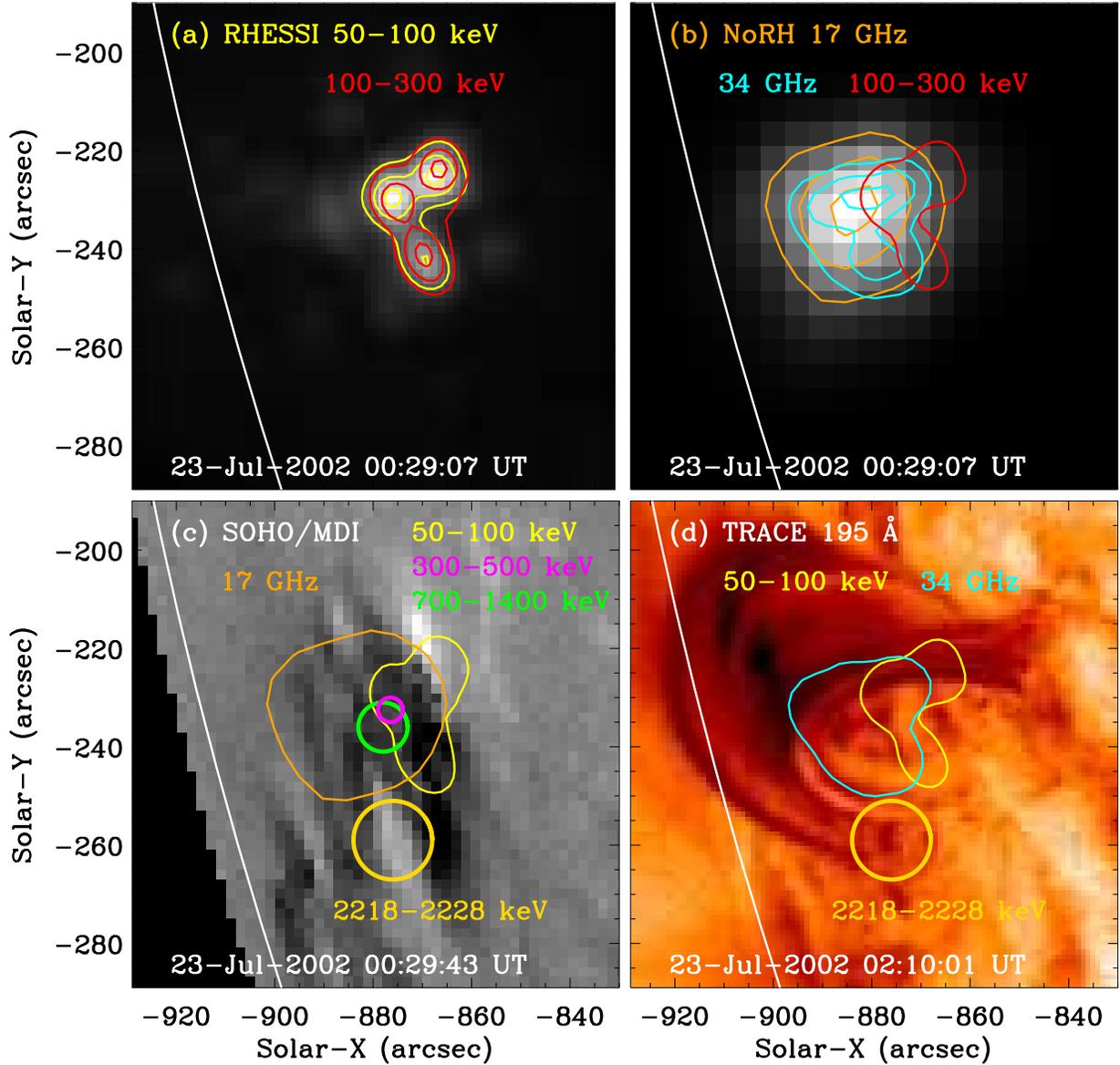}
\caption{Imaging observations of the X4.8 flare with a FOV of
$\sim$100\arcsec$\times$100\arcsec. The background images are
measured by RHESSI~50$-$100~keV (a), NoRH~17~GHz (b), SOHO/MDI (c),
and TRACE~195~{\AA} (d), respectively. The overlaid contours
represents HXR and radio emissions derived from RHESSI 50$-$100~keV
(yellow) and 100$-$300~keV (red), as well as NoRP~17~GHz (orange)
and 34~GHz (cyan). Their levels are set at 30\%, 60\% and 90\%. The
circles represent the source locations
\citep[cf.][]{Hurford03,Lin03} of RHESSI~300$-$500~keV (magenta),
700$-$1400~keV (green), and 2218$-$2228~keV (gold). \label{img}}
\end{figure}

The difference in QPP periods between $\gamma$-ray line and
HXR/radio emissions might be related to their source locations in
the solar surface, as shown in Figure~\ref{img}. Panels~(a) and (b)
present HXR and radio maps with a field-view-of (FOV) of
$\sim$100\arcsec$\times$100\arcsec\ during the X4.8 flare, which is
also in the QPP duration. Here, the HXR maps were reconstructed by
the RHESSI team using the CLEAN algorithm, and we downloaded them
from the RHESSI Image Archive. The overlaid contours were obtained
from the RHESSI data in the energy range of 50$-$100~keV (yellow)
and 100$-$300~keV (red), the NoRP data at frequencies of 17~GHz
(orange) and 34~GHz (cyan). The contour levels are set at 30\%, 60\%
and 90\%, respectively. It can be seen that the two HXR source
locations match well with each other, and their locations are
roughly consistent with the radio source region, which is similar to
previous observations \citep{White03}. The slight divergence could
because that the HXR emission above 50~keV tends to appear in the
double footpoint locations, while the radio emission in 17~GHz is
dominated by the loop-top source, and the radio emission in 34~GHz
moves closely to the footpoint locations.

Figure~\ref{img}~(c) shows the line-of-sight (LOS) magnetogram with
a same FOV observed by SOHO/MDI during the X4.8 flare. Panel~(d)
draws the EUV~195~{\AA} map captured by TRACE after the X1.4 flare,
since the TRACE maps were severe saturated during the intense flare.
The overlaid contours are made from the HXR~50$-$100~keV (yellow)
and radio~17~GHz (orange) or 34~GHz (cyan) emissions at the level of
30\%, similarly to what have been shown in Figure~\ref{img}~(a) and
(b). The overplotted circles represent the source sites of flare
$\gamma$-ray emissions at 2218$-$2228~keV (gold), 700$-$1400~keV
(green), and 300$-$500~keV (magenta), respectively. It
should be pointed out that the $\gamma$-ray source locations are
referenced from \cite{Hurford03} and \cite{Lin03}, because we cannot
restructure the $\gamma$-ray maps of a solar flare. As reported by
\cite{Hurford03}, the centroid locations of $\gamma$-ray continuum
emissions at 700$-$1400~keV and 300$-$500~keV were coincident with
that of the HXR emission at 50$-$100~keV. Moreover, they were both
overlaid on the radio sources at 17~GHz and 34~GHz. On the other
hand, the $\gamma$-ray line centroid position was located away from
the HXR and radio sites, implying that the acceleration region of
energetic ions was displaced from the corresponding site of
nonthermal electrons, i.e., a departure of $\sim$20$\pm$6\arcsec\
\citep[cf.][]{Hurford03,Lin03}. The EUV map after the X4.8 flare
exhibits a series of postflare loops, as shown in
Figure~\ref{img}~(d). Although the $\gamma$-ray line centroid
position was not located near any clear EUV or H$\alpha$ brightening
\citep[cf.][]{Lin03,Vilmer11}, it was actually sited at the
footpoint of a outer and larger postflare loop, while the
$\gamma$-ray continuum centroid locations matched with inner and
smaller loops. All those observations suggest that the propagation
process or acceleration sites of energetic ions and electron beams
might be significantly displaced in the solar flare. The location
differences between $\gamma$-ray line and continuum as well as
HXR/radio emissions are consistent with their variations in
quasi-periods.

\section{Summary and Discussion}
We reported the observation of non-stationary QPPs in the
$\gamma$-ray line emission during the impulsive phase of an X4.8
flare on 2002 July 23. The intense flare was well studied in
the early era of RHESSI, such as: reconstructed maps in $\gamma$-ray
and HXR emissions \citep{Hurford03,Krucker03,Lin03,White03}, Doppler
redshifts and line broadening of heavy particles \citep{Smith03},
spectral measurements in HXR, $\gamma$-ray line and continuum
emissions \citep{Holman03,Murphy03,Share03}. On the other hand, the
quasi-periodicity of this flare was not yet investigated in detail.
In order to simultaneously obtain the high temporal cadence and SN
ratio, we used a broad energy range (i.e., 100~keV) to integrate the
$\gamma$-ray line flux with a time bin of 4~s. In the previous study
\citep[e.g.][]{Hurford03}, a narrow energy range (i.e., 10~keV) but
a low time cadence such as 40~s was applied to improve the SN ratio.
Herein, we could detect the quasi-period of 50$\pm$15~s in the
$\gamma$-ray line emission. Previous observations
\citep[e.g.][]{Chupp83,Nakariakov10,Li20} have showed the presence
of QPP features in $\gamma$-ray light curves during solar flares.
However, these studies covered much broader energy ranges, such as
4100$-$6400~keV \citep{Chupp83} and 2000$-$6000~keV
\citep{Nakariakov10}, or they just based on the $\gamma$-ray
continuum emission at a lower energy range, i.e., 331$-$1253~keV
\citep[e.g.][]{Li20}. To the best of our knowledge, this is
the first detailed study of flare QPPs in $\gamma$-ray line emission
at around 2223~keV.

\begin{figure}
\centering
\includegraphics[width=\linewidth,clip=]{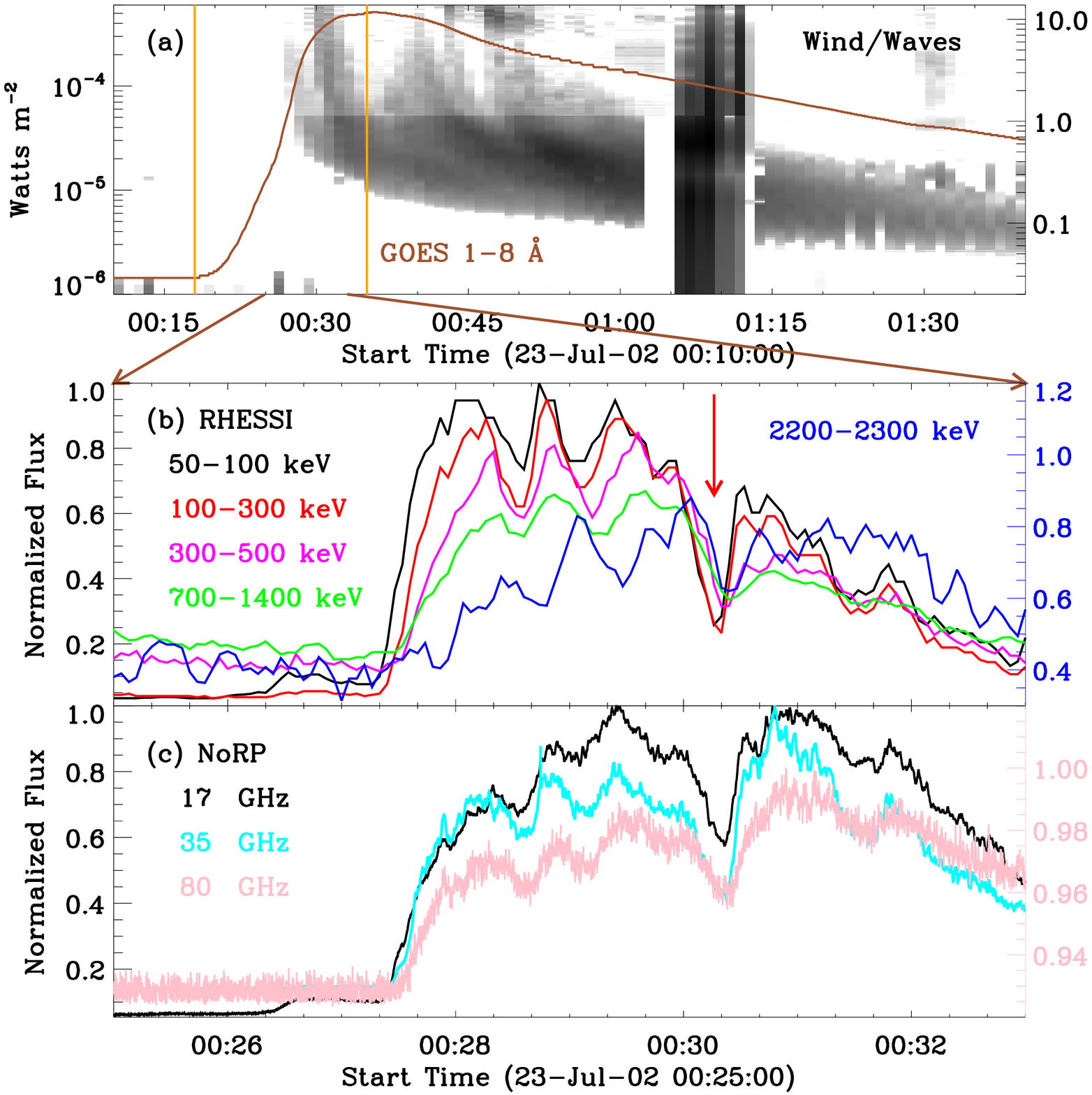}
\caption{(a): Radio dynamic spectrum measured by Wind/Waves. The
overplotted light curve is recorded by GOES~1$-$8~{\AA} from
00:10~UT to 01:40~UT. The vertical orange lines mark the start and
peak times of the X4.8 flare. (b): Normalized HXR and $\gamma$-ray
fluxes between 00:25~UT and 00:33~UT observed by RHESSI in energy
ranges of 50$-$100~keV (black), 100$-$300~keV (red), 300$-$500~keV
(magenta), 700$-$1400~keV (green), and 2200$-$2300~keV (blue). (c):
Normalized radio fluxes during 00:25$-$00:33~UT measured by NoRP at
frequencies of 17~GHz (black), 35~GHz (cyan), 80~GHz (pink).
\label{flux}}
\end{figure}

The flare QPP with a quasi-period of 90$\pm$20~s was observed in
$\gamma$-ray continuum emissions at 300$-$500~keV and 700$-$1400~keV
during the same impulsive phase. The same 90-s QPP was also seen in
the HXR emission at 50$-$100~keV, and radio emissions at frequencies
of 17~GHz and 80~GHz. In fact, the HXR emission at 100$-$300~keV and
radio emission at 35~GHz also showed the similar 90-s QPP, as shown
in Figure~\ref{flux}~(b) and (c). Our observation was consistent
with the observational result from the $\gamma$-ray and HXR maps,
for instance, the centroid locations of $\gamma$-ray continuum
emissions roughly coincided with the source regions of HXR and radio
emissions \citep[cf.][]{Hurford03,White03}. All those observational
facts suggested that the $\gamma$-ray continuum emission at lower
energy was dominated by the electron bremsstrahlung radiation,
similar to the HXR emission \citep{Vilmer11,Vilmer12}. The
modulation depth of the flare QPP was estimated to 20\%$-$30\% in
$\gamma$-ray and HXR emissions, in agreement with what was found by
\cite{Nakariakov10} in microwave, HXR and $\gamma$-ray time series.
The similar modulation depth was also detected in radio emissions at
frequencies of 17~GHz and 35~GHz. However, the modulation depth in
radio~80~GHz was much smaller, i.e., only 1\%$-$2\%, which could be
attributed to the strong background emission and low SN ratio.

The flare QPPs were simultaneously observed in $\gamma$-ray line and
continuum, HXR and radio emissions, implying a common origin of the
flare-accelerated ions and electrons, for instance, they were both
likely accelerated by the repetitive magnetic reconnection during
the flare impulsive phase. Our observations support the idea that
the acceleration process of energetic ions and nonthermal electrons
were linked to each other in the solar flare. On the other hand, the
quasi-period seen in the $\gamma$-ray line emission was visual
shorter than that found in $\gamma$-ray continuum, HXR and radio
emissions, suggesting that there were some variations in
acceleration or propagation process between energetic ions and
electron beams, which had been confirmed by the spatial-resolved
imaging observation, for instance, the neutron-capture line location
was displaced by $\sim$20$\pm$6\arcsec\ from the source region of
electron beams \citep[e.g.][]{Hurford03,Lin03,Vilmer11}.

The flare QPPs can be modulated by several mechanisms, i.e., the
global kink-mode or sausage-mode waves \citep{Nakariakov20}. They
both could cause changes of the magnetic-mirror ratio, and then
drive quasi-periodic precipitations of energetic ions and nonthermal
electrons in flare/coronal loops
\citep[e.g.][]{Foullon05,Nakariakov10}. Hence, their quasi-periods
($P$) are usually associated with loop lengths ($L$) and the
Alfv\'{e}n speed ($V_A$) outside the loop, such as
$P~\approx~2L/V_A$. If we assumed a semi-circular shape for the
flare loop \citep{Tian16}, the short-loop length was estimated to
31~Mm, and the long-loop length was roughly equal to 53~Mm. Thus,
the 90-s QPP observed in HXR and radio emissions in the short flare
loop requires a Alfv\'{e}n speed of $\sim$700~km~s$^{-1}$, while the
50-s QPP seen in the $\gamma$-ray line emission in the long flare
loop needs a Alfv\'{e}n speed of $\sim$2100~km~s$^{-1}$. These
speeds are slower than the typical speed estimated in the global
sausage oscillation, such as 3000$-$5000~km~s$^{-1}$
\citep{Nakariakov03,Melnikov05}, but they are roughly close to the
phase speed of global kink oscillations in flare loops
\citep{Foullon05}. On the other hand, the quasi-period of 50~s seen
in the $\gamma$-ray line emission was associated with a large
postflare loop \citep[e.g.][]{Lin03}, similarly to what was found by
\cite{Dauphin07} that energetic ions were injected in larger loops
than nonthermal electrons. However, our observation contradicted the
previous finding that the quasi-period of kink oscillations was
proportional to the major length of coronal loops
\citep[e.g.][]{Anfinogentov15}. Therefore, there must be effective
mechanisms/models whereby energetic ions were preferentially
accelerated in larger flare loops, such as: the stochastic
acceleration mechanism \citep[e.g.][]{Emslie04}, or the
trap-plus-precipitation model \citep[e.g.][]{Dauphin07}. All those
considerations therefore support a probable interpretation of flare
QPPs: they can be driven by the repetitive magnetic reconnection
that are likely to be triggered by global kink-mode oscillations,
and then produce quasi-periodic particle accelerations, for
instance, periodically accelerated energetic ions and nonthermal
electrons \citep{Foullon05,Nakariakov10}. Of course, the flare QPPs
could be caused by the other repetitive magnetic reconnection, i.e.,
modulated by a self-induced oscillation \citep{Zimovets21a}. So far,
we cannot exclude or demonstrate this possibility, largely due to
the absence of quantitative theories and high-resolution
observations.

\acknowledgments The authors thank the anonymous referee for
his/her valuable comments and suggestions. We thank the teams of
RHESSI, NoRP, NoRH, WIND, TRACE, and SOHO/MDI for their open data
use policy. This work is funded by the National Key R\&D Program of
China 2021YFA1600502 (2021YFA1600500), NSFC under grants 11973092,
U1931138, 12073081, 11820101002, as well as CAS Strategic Pioneer
Program on Space Science, Grant No. XDA15052200, XDA15320300, and
XDA15320301. D. Li is also supported by the Surface Project of
Jiangsu Province (BK20211402).

\end{document}